\documentclass[reprint,amsmath,amssymb,aps,prl,groupedaddress,nofootinbib,twocolumn,superscriptaddress]{revtex4-1}
\usepackage{graphicx}
\usepackage{amsthm,amssymb,amsmath,braket,mathdots}
\usepackage{bm}
\usepackage[pagebackref=false,pdfnewwindow=true]{hyperref} 
\usepackage{epstopdf,psfrag}
\usepackage{relsize,amsbsy}
\usepackage[export]{adjustbox}

\usepackage{graphicx,xcolor,tikz}

\newcommand{\be}{\begin{equation}}
\newcommand{\ee}{\end{equation}}

\newcommand{\bit}{\begin{enumerate}}
	\newcommand{\eit}{\end{enumerate}}

\definecolor{bananayellow}{rgb}{1.0, 0.88, 0.21}
\definecolor{straw}{rgb}{0.32, 0.28, 0.1}

\begin{document}

         \title{Bond disordered spin liquid and the honeycomb iridate H$_3$LiIr$_2$O$_6$ -- \\abundant low energy density of states from random Majorana hopping}
		\author{Johannes Knolle}
		\affiliation{\small Blackett Laboratory, Imperial College London, London SW7 2AZ, United Kingdom}
		\author{Roderich Moessner}
		\affiliation{\small Max-Planck-Institut fur Physik komplexer Systeme, Nothnitzer Str. 38, 01187 Dresden, Germany}
	        \author{Natalia B. Perkins}
		\affiliation{\small School of Physics and Astronomy, University of Minnesota, Minneapolis, Minnesota 55455, USA}
		\date{\today}

		\begin{abstract}
The 	 5d-electron	honeycomb compound H$_3$LiIr$_2$O$_6$  [K. Kitagawa {\it et al.}, Nature 554, 341-345 (2018)]
exhibits an apparent quantum spin liquid (QSL) state. In this  intercalated spin-orbital compound, a remarkable 
pile up of low-energy  states was experimentally observed in specific heat and  nuclear magnetic (NMR) spin relaxation.
We show that a 
		bond disordered Kitaev model can naturally account for this phenomenon, suggesting that 
 disorder plays an essential role in its theoretical description.
 In the exactly soluble Kitaev model, we obtain, via spin fractionalization,  a random bipartite hopping problem of Majorana fermions 
 in a random flux background. This has a divergent  low-energy density of states (DOS) of the required power-law form $N(E)\propto E^{-\nu}$ with a drifting exponent which takes on the value $\nu \approx 1/2$ for relatively strong bond disorder.Breaking time-reversal symmetry removes the divergence of the density of states, as does applying a magnetic field in experiment. We discuss the implication of our scenario, both for future experiments and from a broader perspective.
		\end{abstract}
	
		\maketitle

\textit{Introduction.}-
Recent years have seen remarkable progress in identifying candidate materials that can realize  QSL phases~\cite{anderson1973resonating,Balents2010,knolle2018field}.  
In particular, a significant experimental and theoretical effort has been devoted to the
study of magnetic properties of 4d and 5d systems, such as iridates and ruthenates,   in which  the  interplay  of  
strong spin-orbit coupling and electronic correlations    gives rise to highly anisotropic and spatially dependent  
Ising-like interactions  between effective moments $J_{\text{eff}}=1/2$ \cite{Jackeli2009}.  In  structures with edge-sharing  octahedra, these so-called Kitaev interactions~\cite{Kitaev2006} often dominate over other exchange interactions \cite{Jackeli2009,Chaloupka2010,chun2015direct} which raises the exciting possibility that the corresponding materials
  realizing the  Kitaev QSL  with well defined  fractionalized  excitations ~\cite{Kitaev2006,banerjee2016proximate,Sandilands2015,Nasu2016,Banerjee2017,Do2017,hermanns2018physics}.

 However,  at sufficiently low temperatures almost all  known Kitaev materials order magnetically  rather  than  exhibiting  spin-liquidity~\cite{winter2017models}. A recent exception,  the hydrogen intercalated iridate H$_3$LiIr$_2$O$_6$, remains in a liquid state down to lowest temperatures without any sign of long ranged ordered magnetism~\cite{Kitagawa2018spin}. 
 H$_3$LiIr$_2$O$_6$ is a material belonging to a larger class of iridates on the honeycomb lattice with experimentally confirmed strongly bond-anisotropic exchange interactions~\cite{chun2015direct}.
 H$_3$LiIr$_2$O$_6$ is obtained from the parent  $\alpha$-Li$_2$IrO$_3$ compound by replacing the interlayer Li$^+$ ions 
 with H$^+$ ions and leaving the LiIr$_2$O$_6$ honeycomb plane stoichiometrically unchanged.   
  Because of  weaker interlayer coupling, H$_3$LiIr$_2$O$_6$ compound   is   effectively  quasi two-dimensional material, however with a large number of stacking faults~\cite{Kitagawa2018spin}. For such an intercalated compound, it is expected that defects play an important role because of
the uncontrolled location of  hydrogen ions whose positions strongly
affects the local magnetic interactions due to hybridization with the bridging oxygen ions~\cite{li2018role}.

\begin{figure}
	\centering
	\includegraphics[width=0.8\linewidth]{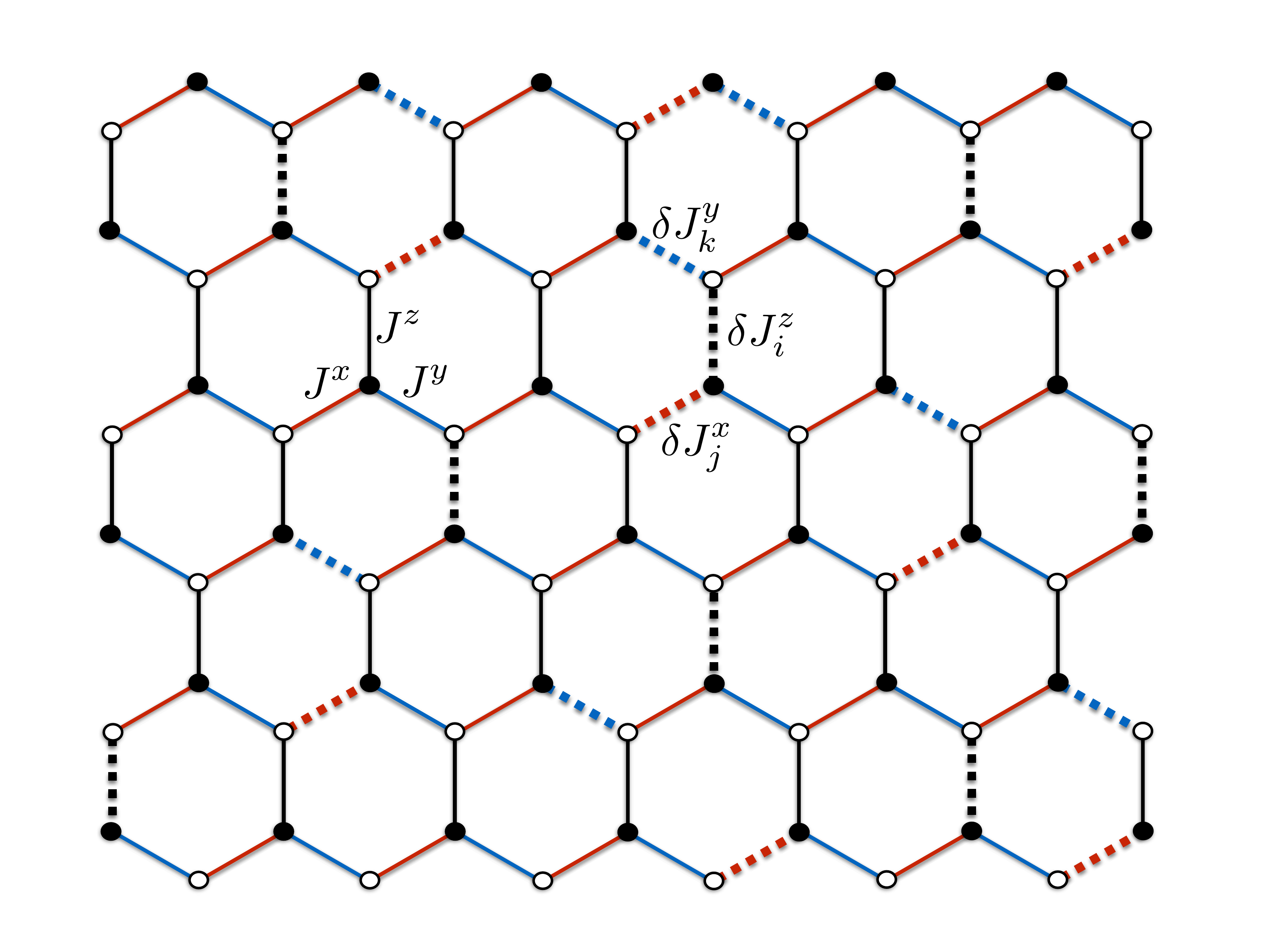}
		\caption{(Color online) The bond-anisotropic Ising interactions of the Kitaev model are shown in three different colours along the three inequivalent bonds. Disorder is taken into account via a subset of randomly selected bonds (dashed lines). }
			\label{Fig1}
\end{figure}
Crucially, the following peculiar features observed in H$_3$LiIr$_2$O$_6$~\cite{Kitagawa2018spin} call for a coherent explanation: i) the specific heat displays a low-temperature divergence of $C/T\propto T^{-1/2}$; ii) only a small fraction of the total magnetic entropy is released at these low-temperature scales;  and iii) there is a non-vanishing contribution down to lowest temperature in the NMR rate $1/T_1$ and an almost flat Knight shift. All of these signal the presence of abundant low-energy DOS related to magnetic excitations. However, despite the presence of dominating Kitaev exchange this phenomenology is at odds with the thermodynamics of the Kitaev QSL~\cite{Nasu2014,Yoshitake2016fractional}, which has a vanishing specific heat and a significant release of half of its total entropy at low $T$. 

So far only few attempts at explaining the new data of H$_3$LiIr$_2$O$_6$ have been put forward.  A proposal by  Slagle {\it et al.}~\cite{Slagle2018theory} is based on a special stacking of a four layer honeycomb structure requiring a considerable degree of fine tuning.  Alternatively, a more
 general proposal  by Kimchi  {\it et al.}~\cite{kimchi2018heat} not based on Kitaev physics connects  the  power-law temperature dependence of the specific heat  in various spin liquids to the formation of disorder-induced networks of local moments in a random singlet phase~\cite{Dasgupta1980,Bhatt1982}. 

Here, we adapt a different viewpoint to account for  the  new observations in H$_3$LiIr$_2$O$_6$, to  reconcile its phenomenology with the {\it presence of disorder and dominating Kitaev exchange}. We show that a pinned random flux background removes the large low-temperature 
entropic contribution and that a particular type  of bond disorder overall accounts for the salient experimental signatures i)-iii). Our scenario is based on the observation that spins in the Kitaev spin liquid fractionalize into itinerant Majorana fermions and localized Z$_2$ gauge fluxes -- equivalent to Majorana fermions hopping on the honeycomb lattice. Crucially, randomness in exchange couplings translates into random hopping strengths.

Our work, complementary to previous studies  of {\it isolated} impurities in the Kitaev model~\cite{Willans2010,Willans2011}, 
 links directly to the well studied random hopping problems in the context of the quantum Hall effect~\cite{gade1993anderson,ludwig1994integer,mudry2003density,Motrunich2001griffiths}. There, in the presence of particle-hole symmetry a divergent low-energy DOS appears naturally and has a  general form $N(E)\propto \frac{1}{E}e^{-\text{const}|\ln E|^{2/3}}$~\cite{Motrunich2001griffiths}, which however does not give the experimentally observed divergence of the specific heat~\cite{Kitagawa2018spin}. Moreover, in the standard bipartite disorder problems, the divergence of the DOS is manifest only at asymptotically small energy scales making it hard to study it even numerically~\cite{Motrunich2002particle}. Thus, it is likely placed outside experimentally accessible regimes at least for magnetic materials.  Nevertheless, strong disorder in the Kitaev model may affect not only the hopping of Majorana fermions but also the flux sector~\cite{Zschocke2015}. 
 We show that our experimentally motivated extension of the random hopping problem to a particular type of binary disorder gives rise to the desired divergence of the DOS and therefore specific heat. 

\begin{figure}
	\centering
	\includegraphics[width=0.9\linewidth]{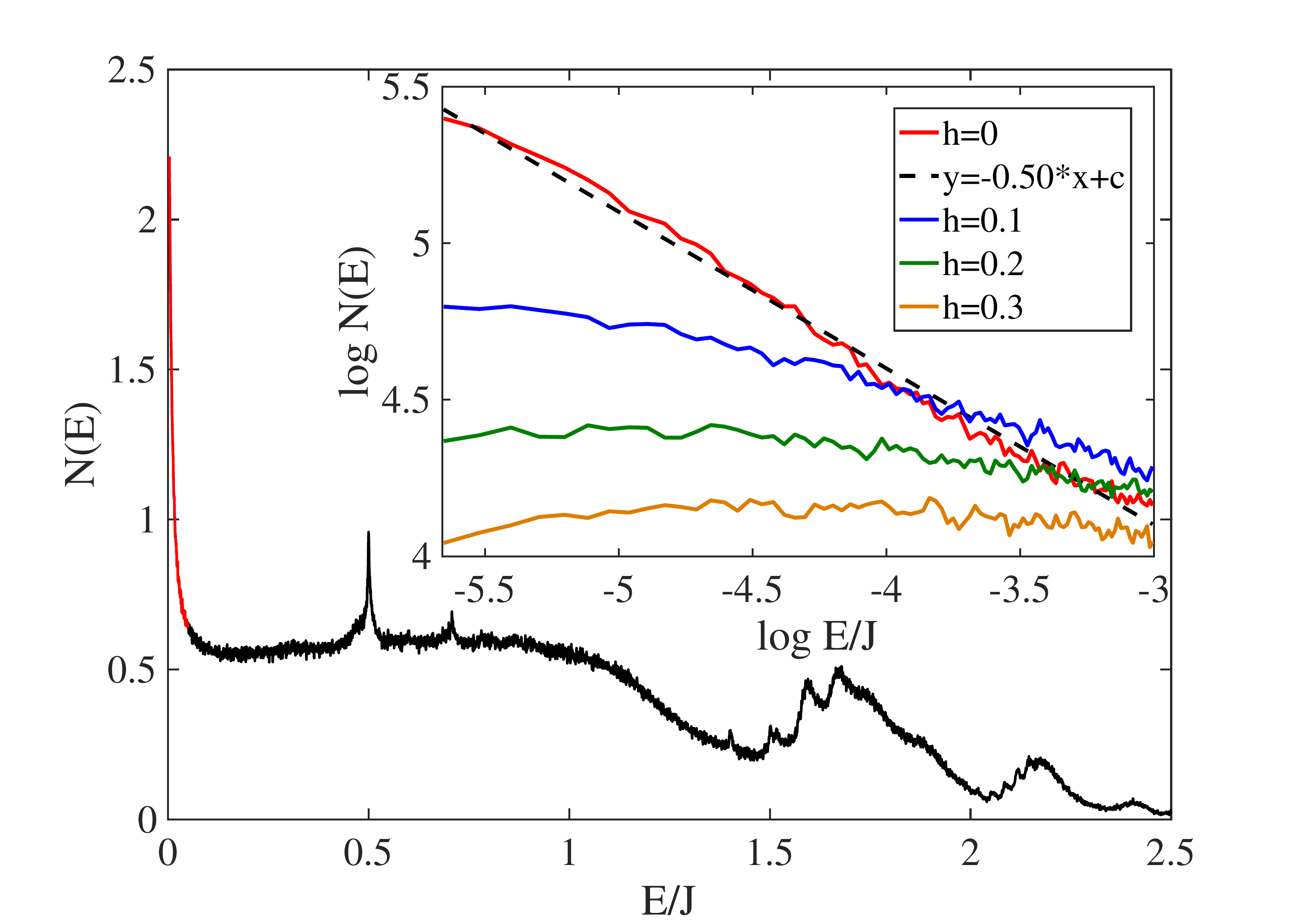}
		\caption{(Color online) The DOS of the random bond, random flux model  Eq.(\ref{HKmodel1}).   For unbroken TRS ($h=0$), $N(E)$  reveals a low-energy divergence. The inset shows the low-energy data (highlighted in red of the main panel) on a log-log plot. Power-law fit (black dashed) of the zero-field  curve ($h=0$ in red) gives  $N(E)\propto E^{-\nu}$ with $\nu= 0.50\pm 0.005$. Breaking TRS by increasing $h$ gradually removes the low-energy divergence. All data are calculated on a torus with $2\times 20^2=800$ spins, 5000 disorder samples for the average and a Lorentzian approximation for the delta function with broadening $\gamma=0.001$.}
	\label{Fig2}
\end{figure}

\textit{Bond disordered Kitaev model.-}
We focus our discussion on a minimal exactly soluble model
\begin{align}
\label{HKmodel1}
H & = & \sum_{\langle i j \rangle_{\alpha}} \left( J^\alpha + \delta J^\alpha_{ij} \right) { \sigma}_i^{\alpha} { \sigma}_j^{\alpha} + h\sum_{\langle \langle i k \rangle \rangle} { \sigma}_i^{\alpha} { \sigma}_j^{\beta} { \sigma}_k^{\gamma} +H_{\text{Flux}},
\end{align}
where  ${ \sigma}^\alpha_i$  being  the Pauli matrices with spin components $\{\alpha,\beta,\gamma\}=x,y,z$ which also label the inequivalent bond directions $\langle i j \rangle_{\alpha}$ on the honeycomb lattice, see Fig.~\ref{Fig1}.   

The first term is the usual Kitaev term with three competing Ising exchanges along the inequivalent bond types, which has been shown to be the leading term of an effective spin $1/2$ description of the LiIr$_2$O$_6$ honeycomb plane~\cite{Singh2012,winter2017models}. Throughout we focus on an isotropic system, $J^{x,y,z}=J>0$, and measure energies in units of $J$. 
The special ingredient is the bond disorder contribution $ \delta J^\alpha_{ij}$ on a random subset of nearest neighbors bonds,  see Fig.~\ref{Fig1}. Along these bonds, spins have an additional \textit{random binary interaction}, $\delta J^{\alpha}_{ij} = \pm \delta J$. More explicitly, we set $\delta J=0.8$ and assume that a density   $\rho=25\%$ of all bonds have a random binary total strength $0.2J$ or $1.8J$  (small variations of $\delta J$ do not alter the results). Note, for $\rho=100\%$ we recover a fully disordered model with random binary bonds.  
The second term in Eq. (\ref{HKmodel1}) is  a three-spin interaction on the three adjacent sites of strength $h$, which
 breaks time-reversal symmetry (TRS). It mimicks the effect of a magnetic field but preserves exact solubility of the model~\cite{Kitaev2006}. 
The last term $H_{\text{Flux}}$ pins a random flux background as further discussed below.  

 Decomposing spin operators into four Majorana fermions~\cite{Kitaev2006}, ${ \sigma}^{\alpha}_i=i c_i b_i^{\alpha}$, the Hamiltonian becomes
\begin{align}
\label{HKmodel2}
H\! & \!= \!\!&\!\!-\!\! \sum_{\langle i j \rangle_{\alpha}} \!\left(J +\delta J^\alpha_{ij} \right) i u_{ij} c_i c_j \! -\! h\!\sum_{\langle \langle i k \rangle \rangle} \!\! i u_{ij} u_{kj} c_i c_k +H_{\text{Flux}}.
\end{align}
Spin excitations fractionalize into Majorana matter fermions $c_i$ and the gauge invariant plaquette fluxes $W_p=\prod_{\langle ij \rangle \in p} u_{ij}$ with the link operators $u_{ij}=i b_i^{\alpha} b_j^{\alpha}$. Both $W_p=\pm 1$ and the gauge dependent link variables $u_{ij}=\pm1$ of Eq.(\ref{HKmodel2}) are constants of motion corresponding to a static background Z$_2$ gauge field. 
While the ground state of the clean Kitaev honeycomb model is flux free, $W_p=+1$ for all plaquettes, the average gap to flux excitations is reduced  for increasing bond disorder~\cite{Knolle2016b}. Here, instead of choosing the particular disordered flux configuration with the lowest ground state energy we assume a
second type of disorder -- we treat Z$_2$ fluxes and hence the $u_{ij}$ as independent random variables with equal probability $u_{ij}=\pm1$.  In a microscopic description this can be easily achieved by adding a chemical potential for flux operators with a random sign,  $H_{\text{Flux}}=\sum_p \mu_p W_p$, with $\mu_p=\pm \mu$ and $\mu\gg J$, which directly freezes a random flux configuration for all relevant temperatures and, therefore, removes the large entropy release of the Kitaev QSL at low temperatures. Alternatively, a  vanishing of the energy scale 
discriminating between flux sectors would translate into a random flux background for nonzero temperatures.

For a given configuration of random link variables $\{ u_{ij}\}$ the Hamiltonian is bilinear in Majorana fermions and can be diagonalized in the standard form
\begin{align}
\label{Hmaj}
H^u& = \sum_n \epsilon^u_n \left (f^\dagger_n f_n-\frac{1}{2} \right ).
\end{align}
Here  $f_n$ are complex matter fermions (superposition of two Majorana operators) which label the eigenmodes  with the fermion energies $\epsilon^u_n $
in a given flux sector.

\begin{figure}
	\centering
	\includegraphics[width=0.9\linewidth]{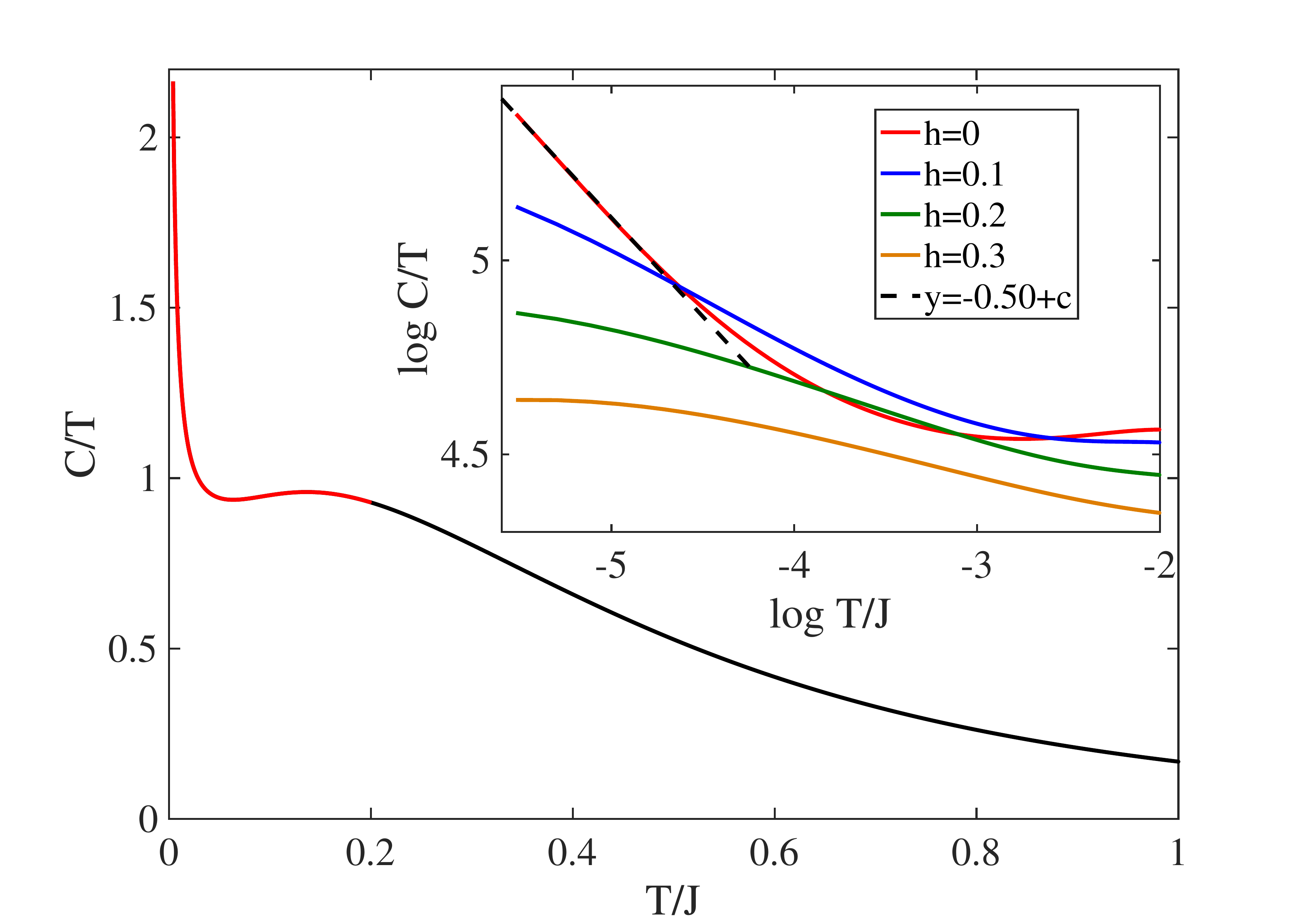}
		\caption{(Color online) Specific heat divided by temperature of the random bond random flux model showing the low temperature divergence. The inset shows the low-$T$ $C/T$ (highlighted in red in the main panel) on a log-log plot. 
Power-law fit (black dashed)   of the zero-field  curve($h=0$ in red) gives $C/T\propto T^{-\nu}$ with $\nu= 0.50$, as expected from the DOS. Breaking TRS by increasing $h$ gradually removes the divergence  as observed in experiments, see  Fig.4(a) in Ref.~\cite{Kitagawa2018spin}.}
	\label{Fig3}
\end{figure}

\textit{DOS and experimental observables.-}
This minimal model, Eq.(\ref{HKmodel2}), displays the salient experimental phenomenology i)-iii). The DOS is calculated as 
\begin{eqnarray}
\label{DOS}
N(E) &= & \left< \left<  \sum_n \delta\left ( E-\epsilon^u_n \right )\right>_{\delta J} \right>_u,
\end{eqnarray}
 the two brackets refer to the independent random bond and random flux averages. 
With this in hand we obtain the specific heat from 
\begin{eqnarray}
\label{SpecificHeat}
C(T) &= & \int \text{d} E  N(E) E \frac{\partial}{\partial T} n_F(E/T)
\end{eqnarray}
with the Fermi function $n_F$ for the thermal occupation of the matter fermions. A divergent DOS of the form $N(E)\propto E^{-\nu}$ translates into a divergent specific heat $C/T \propto T^{-\nu}$. 

The temperature dependence of the NMR spin relaxation rate
 can be obtained from 
\begin{align}
\label{Nuclear}
\frac{1}{T_1} &\propto &    \ \left< \left<  \sum_{\alpha} \mathcal{S}^{\alpha \alpha}_{ii}(\omega\to 0)+\mathcal{S}^{\alpha \alpha}_{ij}(\omega\to 0)  \right>_{\delta J} \right>_u 
\end{align}
with the dynamical on-site ($i,i$) and nearest neighbors ($\langle i,j \rangle$) spin correlations 
\begin{align}
\label{Structure}
  \mathcal{S}^{\alpha \alpha}_{ij}(\omega) =  \int \text{d}t e^{i\omega t}  \langle \sigma^{\alpha}_i(t) \sigma_{ j}^{\alpha}(0) \rangle .
\end{align} 
The brackets correspond to a thermal average at a given temperature $T$.  Since in the Kitaev model only onsite and nearest neighbors correlations are nonzero, the sum in Eq. (\ref{Nuclear}) corresponds to the $\mathbf{q}=0$ component of the spin structure factor. Using the adiabatic approximation of Refs.\cite{Knolle2014,Knolle2015,banerjee2018excitations}, we calculate $\mathcal{S}^{\alpha \alpha}_{\mathbf{q}=0}(\omega)$ via (definition of the flux gap $\Delta$ and transformation matrices $X,Y$ as in \cite{Knolle2015})
 \begin{widetext}
 \begin{eqnarray}
 \label{StructureFactor}
\mathcal{S}^{\alpha\alpha}_{\mathbf{q}=0}(\omega)  & = &   \sum_n \delta \left[\omega-\Delta-\epsilon^u_n \right]\!\! |X_{n,0}|^2 n_F\left(-\epsilon^u_n/T \right) + \sum_n \delta \left[\omega-\Delta+\epsilon^u_n  \right] |Y_{n,0}|^2 n_F\left(\epsilon^u_n/T \right).
 \end{eqnarray}
\end{widetext}
 
Finally, the NMR Knight shift is given by the static susceptibility $\chi$, 
 which we calculate using the Kramers-Kronig relation ($\beta=1/T$)
 \begin{eqnarray}
 \label{KnightShift}
\chi &\propto & \int \text{d}\omega \sum_{\alpha} \mathcal{S}^{\alpha\alpha}_{\mathbf{q}=0}(\omega) \frac{1-e^{\beta \omega}}{\omega} .
 \end{eqnarray}

 \begin{figure}
	\centering
	\includegraphics[width=1.0\linewidth]{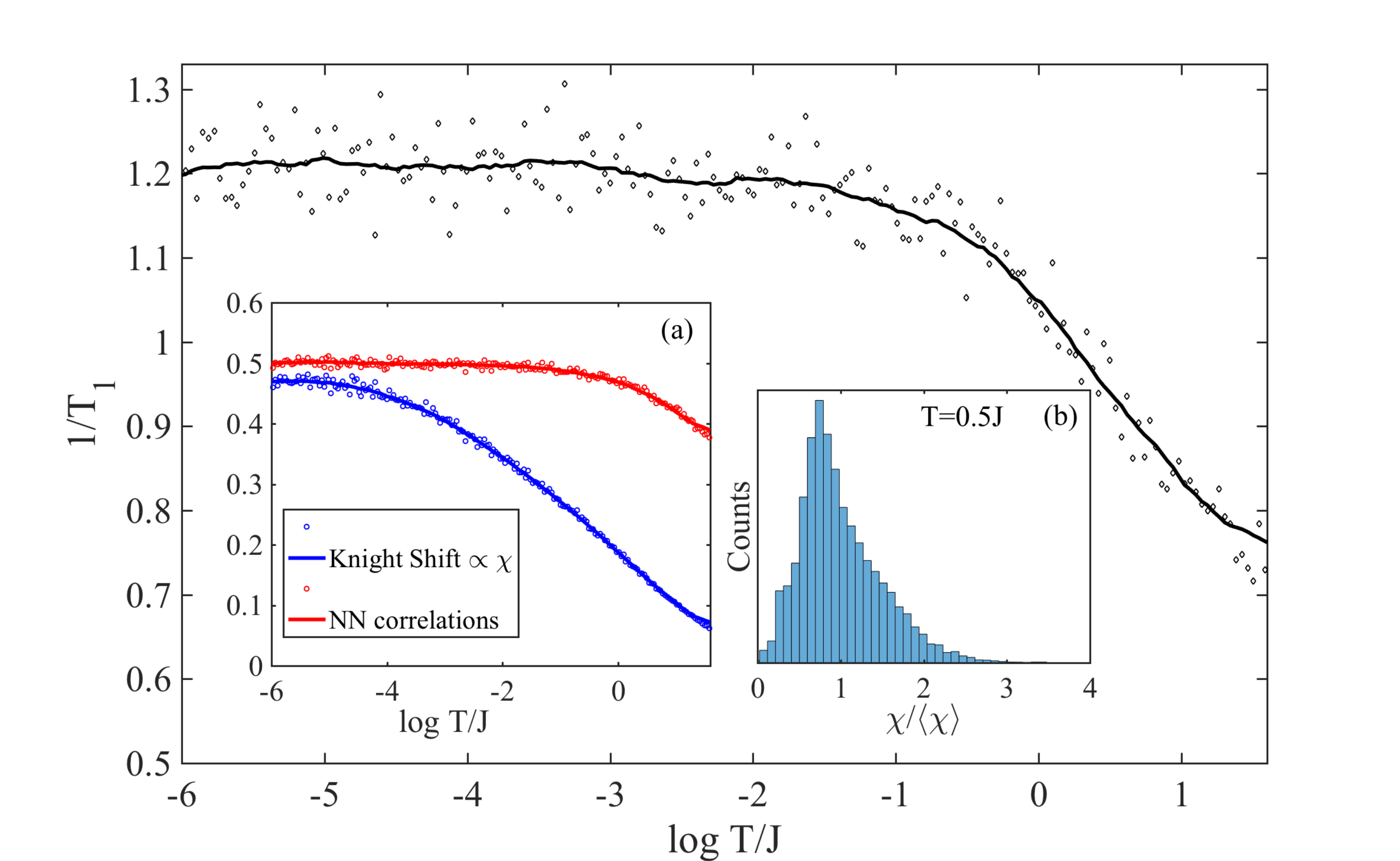}
		\caption{(Color online) Dependence of the NMR relaxation rate $1/T_1$ on temperature for the random bond random flux model with h=0.2. Inset (a) shows the corresponding behavior of the NMR Knight shift (blue) and the static nearest neighbor spin correlations (red). To highlight the qualitative behavior robust to the noise from finite size effects (in lattice size and number of disorder samples) a running average of the data (diamonds) is shown as a solid curve. The distribution of Knight shifts for different disorder configurations (inset (b)) shows a weak {\it relative} shift of a well defined line qualitatively similar to the experiments on H$_3$LiIr$_2$O$_6$, see Fig.2 in Ref.~\cite{Kitagawa2018spin}.}
	\label{Fig4}
\end{figure}

\textit{Results.-} 
In Fig.~\ref{Fig2} we show the DOS which for our choice of parameters, $\rho=25\%$ and $\delta J=0.8$, has a low energy divergence of the required form $N(E)\propto E^{-\frac{1}{2}}$. The resulting divergence of the specific heat is shown in Fig.~\ref{Fig3}. The exponent $\nu=1/2$ does not change for different disorder strength $\delta J$ but is sensitive to the disorder concentration, e.g. it drifts with $\rho$. 
Breaking TRS via a nonzero three spin term gradually 
removes the divergence of the DOS corresponding to the experiments in increasing magnetic fields~\cite{Kitagawa2018spin}. Beyond the low-energy regime, which dictates the thermodynamics at low temperature, 
 the DOS shows the expected spikes of binary disordered systems which originate from special configurations, e.g.  small clusters of strong bonds separated by weak bonds from the rest.

Next, we show the NMR observables in Fig.~\ref{Fig4}. Both the relaxation rate $1/T_1$ (black, main panel) and the nearest neighbors spin correlations (red, inset (a)) are remarkably flat as a function of temperature  showing the presence of 
 low-lying spin excitations down to very low temperature. Only at a temperature scale above the magnetic exchange do they start to decrease towards the paramagnetic regime. Similarly, the Knight shift only slowly decreases for increasing temperature, and its distribution over disorder configurations (inset (b)) shows a small {\it relative} shift of a well defined line. Note, it turns out that because of our choice of symmetric disorder, $\langle \delta J \rangle=0$, the high temperature Curie-Weiss behaviour is insensitive to  disorder strength and concentration. 

\textit{Discussion.-}
Our minimal model is able to reproduce the experimental feature i)-iii) observed in H$_3$LiIr$_2$O$_6$.
We have also considered other forms of disorder, e.g. continuous instead of binary and temperature 
dependent flux densities, however they were not able to give all the low temperature properties observed in experiment. 

A crucial question concerns the microscopic motivation of our particular choice of a binary bond disordered Kitaev model.
 First, similar to other honeycomb iridates~\cite{chun2015direct}, a dominating Kitaev exchange is expected from the intact iridium honeycomb planes~\cite{li2018role,wang2018possible}.  
Second, our particular type of disorder can be generated by
 variations of  the local crystal field due to the random location of the H$^+$  ions  above/below the Ir-Ir bonds which are  naturally   coupled to nearby O$^{2-}$ ions and thus can introduce local distortions of the  oxygen   octahedral cage. The deviation  in crystal fields, in turn, can strongly modify the coupling between Ir ions~\cite{Sizyuk2014,li2018role,wang2018possible}. 
Third, the random flux background is  motivated from the experimental observation that the low temperature release of entropy from the specific heat divergence in H$_3$LiIr$_2$O$_6$ is only a few percent of the total entropy in contrast to the clean Kitaev QSL~\cite{Yoshitake2016fractional}. 
While our example of $H_{\text{Flux}}$ indeed
stabilizes the random fluxes, its microscopic origin is ultimately dictated by the additional interactions beyond the exactly soluble terms considered here, and by the particular disorder physics in real materials. 
 
An important experimental observation  is the scaling of the specific heat divergence as a function of applied magnetic field~\cite{Kitagawa2018spin}. Unfortunately within our exactly soluble model we cannot capture this physics. While in the  pure Kitaev model $h\propto B^3$~\cite{Kitaev2006},  in  the presence of off-diagonal exchange, $\Gamma$, it can be  of the form  $h\propto \Gamma^2 B$~\cite{Song2016} or has a more complicated dependence in real materials.
Nevertheless, we do capture the overall qualitative trend that breaking TRS leads to a suppression of the $C/T$ divergence, see Fig.~\ref{Fig3} inset. 

All our results were calculated for AFM Kitaev interactions but there still remains a considerable debate concerning the overall sign of the exchange in different Kitaev candidate materials~\cite{winter2017models}. An unexpected aspect of the Kitaev QSL -- and arguably in itself a signature thereof -- is the insensitivity of the DOS and specific heat to the overall sign of Kitaev exchange $J$ (even for nonzero $h$ the sign change for the resulting bipartite Majorana hopping model can be removed by a simple gauge transformation). Only the the sign of the nearest neighbour spin correlator  $\mathcal{S}^{\alpha \alpha}_{\langle i,j\rangle}$ depends on the sign of the Kitaev interaction, i.e. the equal time component is negative (positive) for AFM (FM) exchange. However, it turns out that since its magnitude is much smaller than that of the on-site component, there are only small quantitive changes and the qualitative features of the NMR response remain unchanged. Note, the slight difference in temperature dependences for $1/T_1$ measured on different sites in H$_3$LiIr$_2$O$_6$~\cite{Kitagawa2018spin} potentially points to an AFM interaction similarly to the by now classic interpretation in terms of different form factors of NMR lines in copper high-$T_c$ materials~\cite{mila1989analysis}. 

In order to make quantitive contact to the experimental findings in H$_3$LiIr$_2$O$_6$~\cite{Kitagawa2018spin}, a  full microscopic derivation of our minimal scenario is needed, which is an important as well as challenging direction for future research. 
A first step would be an investigation of the phenomenology close to the integrable limit considered here~\cite{Knolle2018dynamics}. 
In the context of random hopping models, it would be interesting to explore the localization properties of the Majorana wave functions and a careful analysis of extra logarithmic corrections to our low energy divergence of the DOS. 

\textit{Summary and Outlook:}
We have argued that a minimal model of a bond disordered Kitaev QSL can account for the salient experimental observations in H$_3$LiIr$_2$O$_6$~\cite{Kitagawa2018spin}, namely i) a low temperature divergence of the specific heat $C/T \propto T^{-\nu}$ with $\nu=1/2$, which ii) carries only a small fraction of the total entropy; and iii) an NMR response signalling an abundant low energy DOS of magnetic excitations. Our scenario directly leads  to a number of qualitative experimental predictions. Firstly, Raman or inelastic neutron scattering experiments should pick up the low energy divergent power-law tail of the DOS. Secondly,  a controlled, sufficiently large change of the disorder concentration should be observable as a drift in the exponent of the divergence. More exotically, in a disordered system of Majorana fermions the breaking of TRS can in principle lead to a thermal metal state with extended wave-functions resulting in a longitudinal thermal transport which diverges logarithmically as a function of sample size~\cite{Mildenberger2007density}. 
More broadly, the experiment in question refocuses attention on the physics of disorder in quantum spin liquids, which has been with
the field for a long time~\cite{villain1979insulating,obradors1988magnetic}, and in particular, which kind of new phenomena may be encountered as 
a cooperative manifestation of disordered topological condensed matter systems~\cite{andreanov2010spin,savary2017disorder,sen2015topological,kimchi2017valence}. 

\textit{Acknowldegements:} We thank P. Mendels for helpful discussions. This work was in part supported by the Deutsche Forschungsgemeinschaft under grant SFB 1143. NP  acknowledges the support from NSF DMR-1511768 Grant.


\bibliography{KitaevDis_refs}


%
%

\end{document}